\documentclass[aip,reprint]{revtex4-1}
\usepackage{amsmath}
\usepackage{amsfonts}
\usepackage{amssymb}
\usepackage{natbib}
\usepackage{graphicx}
\usepackage{setspace}
\usepackage{color}
\usepackage{float}
\usepackage{xcolor}
\usepackage{ulem}
\usepackage{tabularx}
\usepackage{hyperref}
\usepackage[toc,page]{appendix}
\def \rs#1 {{\bf #1}}
\def \tf#1 {#1}

\def\gm{\gamma}
\def\lmb{\lambda}
\def\omg{\omega}

\begin{document}

% Use the \preprint command to place your local institutional report number 
% on the title page in preprint mode.
% Multiple \preprint commands are allowed.
\preprint{}

\title{Revisiting the Strong Shock Problem:\\ Converging and Diverging Shocks in Different Geometries} %Title of paper

% repeat the \author .. \affiliation  etc. as needed
% \email, \thanks, \homepage, \altaffiliation all apply to the current author.
% Explanatory text should go in the []'s, 
% actual e-mail address or url should go in the {}'s for \email and \homepage.
% Please use the appropriate macro for the type of information

% \affiliation command applies to all authors since the last \affiliation command. 
% The \affiliation command should follow the other information.

\author{Elisha Modelevsky}
\email[]{elisha.modelevsky@mail.huji.ac.il}
%\homepage[]{Your web page}
%\thanks{}
\affiliation{Racah Institute of Physics, Hebrew University, Jerusalem 91904, Israel}
\affiliation{Israel Atomic Energy Commission, P.O. Box 7061, Tel Aviv 61070, Israel}

\author{Re'em Sari}
\affiliation{Racah Institute of Physics, Hebrew University, Jerusalem 91904, Israel}

% Collaboration name, if desired (requires use of superscriptaddress option in \documentclass). 
% \noaffiliation is required (may also be used with the \author command).
%\collaboration{}
%\noaffiliation

\date{\today}

\begin{abstract}
    Self-similar solutions to converging (implosions) and diverging (explosions) shocks have been studied before, in planar, cylindrical or spherical symmetry. Here we offer a unified treatment of these apparently disconnected problems . We study the flow of an ideal gas with adiabatic index $\gamma$ with initial density $\rho\sim r^{-\omega}$, containing a strong shock wave. We characterize the self-similar solutions in the entirety of the parameter space $\gamma,\omega$, and draw the connections between the different geometries. We find that only type II self-similar solutions are valid in converging shocks, and that in some cases, a converging shock might not create a reflected shock after its convergence. Finally, we derive analytical approximations for the similarity exponent in the entirety of parameter space.
\end{abstract}

\maketitle
\newpage

This article may be downloaded for personal use only. Any other use requires prior permission of the author and AIP Publishing. This article appeared in Physics of Fluids 33, 056105 (2021) and may be found at \url{https://doi.org/10.1063/5.0047518}.

\section{Introduction}

The strong explosion problem in non-uniform density distribution has been studied extensively. The roots of this problem lie with the work on the strong spherical explosion problem in a uniform density distribution (the so-called Sedov-Taylor problem) \citep{sedov1946,taylor1950,von_neumann1947}, which was solved analytically by Sedov for $\omega<3$. In the works of Waxman \& Shvarts \cite{waxman1993,waxman2010}, it was shown that Sedov's solutions are no longer valid for $\omega>3$ and the correct asymptotic self-similar solutions for $\omega>\omega_g>3$ were obtained in a semi-analytical way. Gruzinov \citep{gruzinov2003} ``closed the gap" in the region $3<\omega<\omega_g$ and thus the strong spherical explosion's self-similar solutions were described for all values of $\omega$. It will be shown in this article that these results can be easily modified to apply for cylindrical geometry, and that for planar geometry a gap does not exist
\citep{kushnir_katz2014}.

The strong implosion problem is the explosion's problem counterpart for a converging shock. This problem can be described as finding a self-similar solution for the flow containing a strong shock that converges towards the origin. This problem, including a description of the shock reflected from the origin, was first solved in uniform density by Guderley \citep{guderley1942}. Sakurai studied the problem of a planar converging shock in power-law density \citep{sakurai1960}. Several authors have studied converging shocks in power-law densities distribution $\rho\sim r^{-\omega}$ and in different geometries \citep{sharma1995,toque2001,madhumita2003}. A thorough study of the solution's dependence on $\omg$ has been conducted recently \citep{giron2021A,giron2021B} for intermediate $\omg$ values.

Diverging and converging strong shocks appear in a variety of physical phenomena. Spherical diverging shocks were first studied to describe the hydrodynamic effect of powerful earthly detonations \citep{von_neumann1947}, and later to analyze the dynamics of supernovae \citep{chevalier1976}. More recently, planar diverging shock solutions were applied to describe stellar collisions \citep{kushnir_katz2014}. Spherical converging shocks appear in physical phenomena such as sonoluminescent bubble collapse\citep{evans1996} and inertial confinement fusion (ICF)\citep{rygg2008,bhagatawala2012}. Planar converging shocks can be used to describe a shock wave approaching the edge of a star from inside\citep{sakurai1960}.

The generalized problem we address in this article can be posed in the following way - find a self-similar solution to the one-dimensional flow equations for an ideal gas with adiabatic index $\gamma$, with a given initial density distribution $\rho_{0}(r)=Kr^{-\omega}$, that includes a strong shock wave propagating either inwards or outwards, in an arbitrary geometry. The geometry is denoted by the logical parameter $n$, with $n=0$ for planar geometry, $n=1$ for cylindrical geometry and $n=2$ for spherical geometry. The direction of the strong shock's propagation is denoted by the logical parameter $s$, with $s=1$ for diverging shocks and $s=-1$ for converging shocks. The dimensionless parameters in this problem are therefore $s,n,\gamma,\omega$.

Assuming the flow is self-similar, it should be possible to describe it using a single dimensionless coordinate $\xi=\frac{r}{R(t)}$, where $r$ and $t$ are spatial and temporal coordinates, and $R(t)$ is the shock front's distance from the origin as a function of time. Throughout this article, $t=0$ will always denote the special time in the flow - the time at which the shock is either at the origin or at infinity, depending on the specific case. Self-similarity implies that $R(t)=At^\alpha$, but this can be problematic in cases where there is either shock convergence or shock divergence in finite time, since $t$ can be both positive and negative. A more easily generalized approach would be to define the similarity coordinate $x=\frac{t}{(r/A)^\lambda}$, where $\lambda=\alpha^{-1}$. When a strong shock exists, its location is at $|x|=1$. From now on, let us assume that units for $r,t$ have been chosen such that $A=1$ and the similarity coordinate is
\begin{equation}
    x=\frac{t}{r^{\lambda}}
\end{equation}

In the special case $\lambda=0$, the similarity coordinate is different. Since this constitutes a set of measure zero in parameter space, we will neglect this case. Table~\ref{table:x_regions} summarizes how the sign of $\lambda$ and the direction of shock propagation determine the solution's dependence on $x$.

\begin{table*}
\label{table:x_regions}
\caption{Partition of the $x$ axis to regions according to $s$ and $\lmb$.}
    \begin{tabularx}{1\textwidth}{| >{\centering\arraybackslash}X 
  | >{\centering\arraybackslash}X | >{\centering\arraybackslash}X
  | >{\centering\arraybackslash}X | >{\centering\arraybackslash}X |}
        \hline
         & \multicolumn{2}{c|}{Converging $(s=-1)$}  & \multicolumn{2}{c|}{Diverging $(s=1)$} \\
         \hline
         & $\lambda>0$ & $\lambda<0$ & $\lambda>0$ & $\lambda<0$ \\
         \hline
        $x<-1$ & Unperturbed fluid & - & - & Unperturbed fluid \\
        \hline
        $x=-1$ & Main shock & - & - & Main shock\\
        \hline
        $-1<x<0$ & Flow behind the shock & - & - & Flow behind the shock\\
        \hline
        $x=0$ & Entire fluid at $t=0$ & - & - & Entire fluid at $t=0$\\
        \hline
        $0<x<B$ & Once-shocked flow & Unperturbed fluid & Unperturbed fluid & Shock-free flow\\
        \hline
        $x=B$ & Reflected shock & The shock & The shock & Shock-free flow\\
        \hline
        $B<x$ & Doubly-shocked flow & Flow behind the shock & Flow behind the shock & Shock-free flow\\
        \hline
    \end{tabularx}
\end{table*}
For infinite-time convergence and divergence, $B=1$. In the case of finite-time convergence, determining the location of the reflected $B$ requires solving an eigenvalue problem as will be described later. As will be seen, $B$ might be infinite in some of parameter space. In the case of finite-time divergence, the value of $B$ is meaningless, since there is no returned shock from infinity. Equivalently, it can be said that $B$ is always inifinite in this case. It can be seen from the table, that in all cases, the strong shock point is at $x=\text{sign}(\lambda s)$.

Let us define the similarity flow functions for the density, material velocity and sound speed:
\begin{equation}
    \label{eq:ansatz}
    \begin{gathered}
        \rho(r,t)=Kr^{-\omega}G(x)\\
        u(r,t)=\lambda^{-1}t^{-1}rU(x)=\lambda^{-1}r^{1-\lambda}\frac{U(x)}{x}\\
        c(r,t)=s\lambda^{-1}t^{-1}rC(x)=s\lambda^{-1}r^{1-\lambda}\frac{C(x)}{x}
    \end{gathered}
\end{equation}
Note that $G$ is the compression ratio of a fluid element in relation to the initial density at its location, and not in relation to the element's initial density. The sound speed $c$ is always non-negative, and the material velocity has $\text{sign}(u)=\text{sign}(s)$ right behind the strong shock. As a result of these definition of the similarity functions, the strong shock point has $U,C>0$ in all cases.

\section{The self-similar solution}
\label{sec:self_similar_solution}
Starting with the spherically-symmetric Eulerian flow equations in terms of $\rho,u,c$
\begin{equation}
    \label{eq:1dflow}
    \begin{gathered}
        \frac{\partial\rho}{\partial t}+\frac{\partial}{\partial r}\left(\rho u\right)+\frac{n\rho u}{r}=0\\
        \frac{\partial u}{\partial t}+u\frac{\partial u}{\partial r}+\frac{1}{\gamma \rho}\frac{\partial}{\partial r}\left(\rho c^2\right)=0\\
        \frac{\partial c}{\partial t}+u\frac{\partial c}{\partial r}+\frac{\gamma-1}{2}c\left(\frac{\partial u}{\partial r}+\frac{nu}{r}\right)=0
    \end{gathered}
\end{equation}

by plugging in the similarity ansatz (\ref{eq:ansatz}) we can turn the PDE system into an ODE system for $U,C$ as functions of $x$.
\begin{equation}
    \label{eq:main_ODE}
    \begin{gathered}
        \lambda x\frac{dU}{dx}=\frac{\Delta_1(U,C)}{\Delta(U,C)}\\
        \lambda x\frac{dC}{dx}=\frac{\Delta_2(U,C)}{\Delta(U,C)}
    \end{gathered}
\end{equation}

where
%\small
\begin{equation}
    \label{eq:Deltas}
    \begin{gathered}
        \Delta(U,C)=(1-U)^2-C^2\\
        \Delta_1(U,C)= \left[\frac{\omega+2(\lambda-1)}{\gamma}-(n+1)U\right] C^2\\+U(1-U)(\lambda-U)\\
        \Delta_2(U,C)= \Bigg[(1-U)^2-\frac{n(\gamma-1)}{2}U(1-U)\\+(\lambda-1)\left(\frac{\gamma-3}{2}U+1\right)\Bigg]C\\
        -\left[1+\frac{2(\lambda-1)-(\gamma-1)\omega}{2\gamma(1-U)}\right]C^3
    \end{gathered}
\end{equation}
%\normalsize
From (\ref{eq:main_ODE}) it is evident that the ODE can be solved for $U,C$ independently of $x$:
\begin{equation}
    \label{eq:dudC}
    \frac{dU}{dC}=\frac{\Delta_1(U,C,\lambda)}{\Delta_2(U,C,\lambda)}
\end{equation}
Equation (\ref{eq:main_ODE}) is the same as in Waxman and Shvarts \citep{waxman1993}, up to a factor of $-\lambda$ resulting from the different self-similar coordinate, and equation (\ref{eq:dudC}) is precisely the same. Note that since $s^2=1$, the equations are independent of $s$, and are thus identical for converging and diverging shocks.

We solve the ODE in the from given by eq. (\ref{eq:main_ODE}).
For completeness, we need to write an equation for the compression function $G$. Fortunately, such an equation can be integrated from (\ref{eq:1dflow}) and we have an algebraic equation for $G$:
\begin{equation}
    \begin{gathered}
    \label{eq:G}
    G^q\left(\frac{C}{x}\right)^2 (1-U)^{q+\gamma-1}=\mathrm{const}\\
    q=\frac{2(1-\lambda)+(n+1)(\gamma-1)}{\omega-(n+1)}
    \end{gathered}
\end{equation}
Of course, this constant is changed by a shock, like the strong shock or the reflected shock. Its value behind the strong shock is given by the shock jump conditions. From the Rankine-Hugoniot equations for a strong shock \cite{landau_lifshitz_fluid_mechanics},
\begin{equation}
\begin{gathered}
\label{eq:strong_shock_point}
    G(x_i)=\frac{\gamma+1}{\gamma-1},\;
    U(x_i)=\frac{2}{\gamma+1},\\
    C(x_i)=\frac{\sqrt{2\gamma(\gamma-1)}}{\gamma+1}
\end{gathered}
\end{equation}
These values are used as the initial conditions at $x_i=\text{sign}(\lambda s)=\pm 1$ for the integration on (\ref{eq:main_ODE}). The value of the similarity exponent $\lambda$ is determined in different ways, according to the region in parameter space.

\subsection{Type I self-similar solutions}
For self-similar solutions of the first kind, the similarity exponent $\lambda$ can be obtained from dimensional considerations, assuming that the total mechanical energy in the solution is conserved. This yields the value
\begin{equation}
    \label{eq:st_exponent}
    \lambda=\frac{n+3-\omega}{2}.
\end{equation}
% In type I self-similar solutions, the energy contained in any interval $x\in[x_1,x_2]$ is conserved. One can examine a surface at $r(t)$ with a constant $x$ (which implies $r\sim t^{1/\lmb}$) and write the energy flux through it, demanding that it vanishes. The energy going through the surface is comprised of two contributions - the energy transferred by the fluid's motion and the energy transferred due to the surface movement. During a time interval $dt$, the fluid carries an energy of $dt\cdot \rho r^n u \left(h+u^2/2\right)$, where $h=\frac{c^2}{\gm-1}$ is the specific enthalpy of the fluid. The energy transferred due to the surface's motion in this time interval is $dt\cdot \rho r^n v_s \left(e+u^2/2\right)$, where $e=\frac{c^2}{\gm(\gm-1)}$ is the specific energy of the fluid and $v_s=\frac{r}{\lmb t}$ is the surface's velocity. Equating these terms:
The energy flux through a surface with constant $x$ must be zero, therefore
\begin{equation*}
	u\left(\frac{c^2}{\gm-1}+\frac{u^2}{2}\right) = \frac{r}{\lmb t} \left(\frac{c^2}{\gm(\gm-1)}+\frac{u^2}{2}\right).
\end{equation*}
Plugging in the self-similar functions (\ref{eq:ansatz}) gives the relation
\begin{equation}
\label{eq:type1_CU}
	C^2=\frac{\gm(\gm-1)(1-U)U^2}{2(\gm U -1)}.
\end{equation}
This curve passes through the strong shock point (\ref{eq:strong_shock_point}). Hence, the solution might go in one of the two branches that extend from it. The first branch extends towards $(U,C)=\left(\frac{1}{\gm},\infty\right)$ and the second branch ends on $(U,C)=(1,0)$. The endpoint of the second branch is a point of contact with vacuum, since it has zero pressure ($C=0$) and the material there maintains a constant value of $x$ ($U=1$). The correct branch can be determined by examining the sign of $\frac{dU}{dx}=\frac{\Delta_1}{\Delta}$ from (\ref{eq:main_ODE}) in the strong shock point, and it can be shown that when $\omega<\omega_h$ the first branch that extends to infinite $C$ applies, and when $\omega>\omega_h$ the second branch that tends towards the vacuum applies.
\begin{equation}
    \label{eq:omega_h}
    \omg_h=1+\frac{3-\gm}{\gm+1}n
\end{equation}
When $\omg=\omg_h$, the entirety of the solution is in the strong shock point. $\omg_h$ is bounded by $1-n\le\omg_h\le1+n$ for $\gamma>1$.

\subsection{Type II self-similar solutions}

In self-similar solutions of the second kind, the $U-C$ curve of the solution must pass through the sonic line \citep{guderley1942} defined by $\Delta=(1-U)^2-C^2=0$ . In order to pass smoothly through the sonic line, the solution must pass through a singular point on the sonic line, that satisfies 
\begin{equation}
    \Delta=\Delta_1=\Delta_2=0.
\end{equation}
This constraint determines the value of the similarity exponent $\lambda$. For planar geometry $n=0$, a single singular sonic point exists, and for cylindrical and spherical geometries $n>0$ there are two such points:
\begin{equation}
\label{eq:sonic_point}
    C_{sonic}=\begin{cases}
    \frac{\gm(1-\lmb)}{\omg+(\gm-2)(1-\lmb)} & n=0\\
    h\pm\sqrt{h^2+\frac{1-\lmb}{n}} & n>1
    \end{cases},
\end{equation}
where $h=\frac{1}{2}-\frac{\omg+(\gm-2)(1-\lmb)}{2n\gm}$, and $1-U_{sonic}=C_{sonic}$. For $n>0$, the solution can pass through only one of the singular sonic points, and the choice of sign in (\ref{eq:sonic_point}) varies for different regions in parameter space. For very large positive values of $\omega$, the positive sign is chosen; for very large negative values of $\omega$, the negative sign is chosen. The choice of sign changes at the point where the two singular points merge.

Once the value of $\lmb$ that makes the solution pass through the singular sonic point is calculated, the integration of (\ref{eq:main_ODE}) can be continued towards $x\to\infty$. In cases where $x_i=\text{sign}(\lmb s)=-1$, the integration will pass through the origin $(U,C)=(0,0)$ at $x=0$. The solution can pass through the origin smoothly \citep{lazarus1981} thanks to the choice of the similarity variable $x=t/r^\lmb$ instead of the more common $\xi=r/t^\alpha$. $x=0$ represents the entirety of the fluid at the convergence/divergence time $t=0$, and positive values of $x$ correspond to $t>0$.

\subsubsection{Reflected shocks}
\label{sec:reflected_shocks}

In some cases, (\ref{eq:main_ODE}) cannot be integrated to $x\to\infty$ since the solution approaches the second singular line $U-C=1$. This indicated the existence of a reflected shock. When a reflected shock exists, which can happen only in converging shocks $s=-1$ and when $\omg$ is less than some $\omg_r(n,\gm)$, there is a certain point $x=B>0$ where the solution exhibits a shock discontinuity.

A method for obtaining the reflected shock solution for $\omg=0$ was developed by Lazarus \citep{lazarus1981}, and recently generalized to arbitrary $\omg$ by Giron et al \citep{giron2021B}. For completeness, we briefly present this method here. First, the $U-C$ curve of the doubly-shocked flow is calculated by switching to a new variable $y=kx^{-\sigma}$ and expanding $U$ and $C$ around $y=0$ ($x=\infty$).
\begin{equation}
\begin{gathered}
U(y)=U_0+U_1 y+... \\
C(y)=-y^{-1}+C_0+...
\end{gathered}
\end{equation}
Plugging these into (\ref{eq:main_ODE}) and equating coefficients of powers of $y$ yields
\begin{equation}
\begin{gathered}
U_0=\frac{2(\lmb-1)+\omg}{(n+1)\gm} \\
\sigma=\frac{1}{\lmb}\left(1+\frac{\lmb-1-\frac{\gm-1}{2}\omg}{\gm(1-U_0)}\right).
\end{gathered}
\end{equation}
This allows the calculation of the $U-C$ curve by starting the integration at some small value of $y_0$ and $U(y_0)=U_0,C(y_0)=y_0^{-1}$. This integration cannot give $U$ and $C$ in terms of $x$ since $k$ is yet to be determined.

\begin{figure*}
	%\centering
	\includegraphics[width=\textwidth]{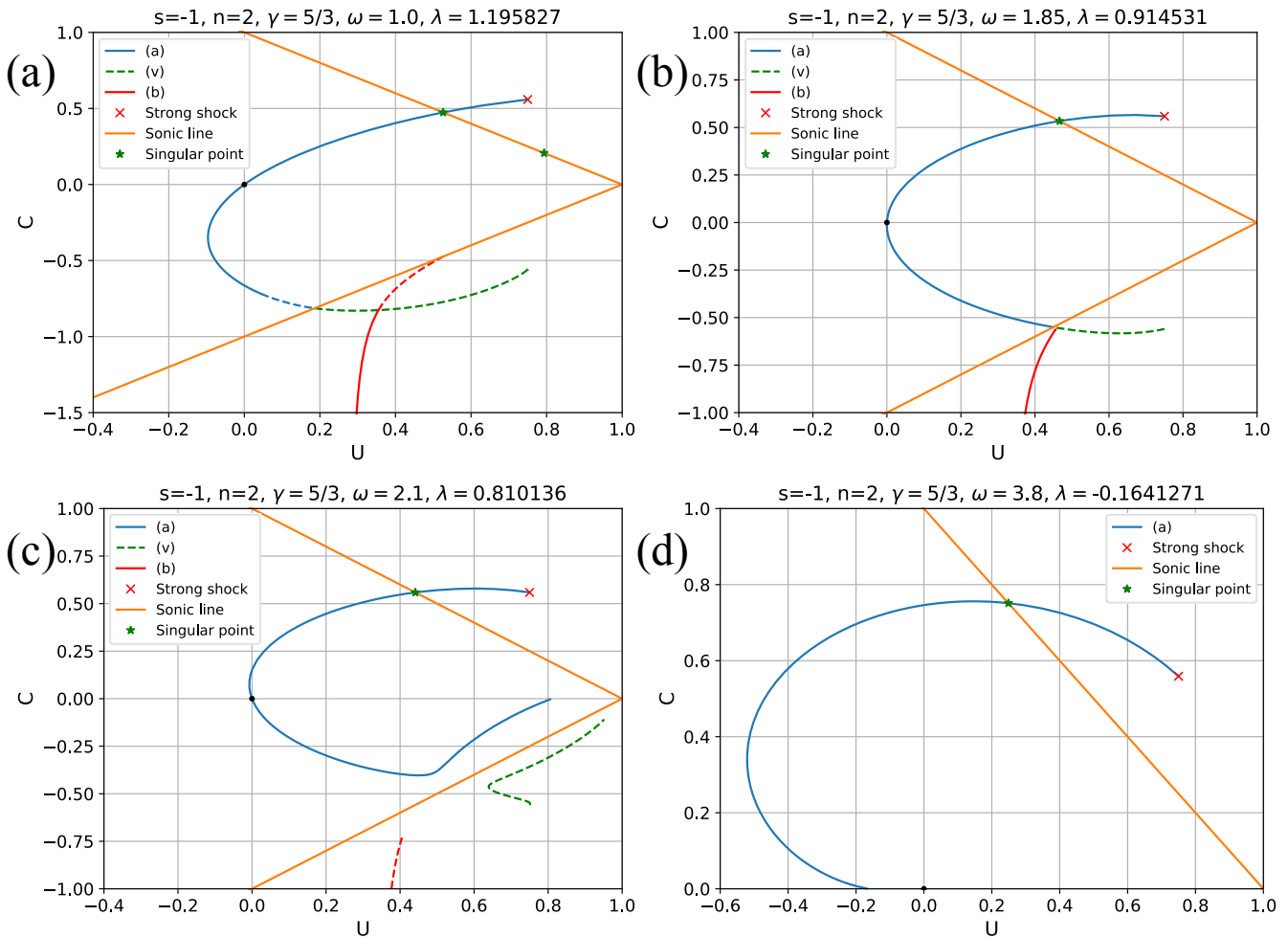}
	\caption{$U-C$ diagrams of solutions to spherical converging shocks with $\gamma=5/3$ and various $\omega$. The blue curve (a) is the solution behind the main shock, starting from the strong shock point (red cross), and going through the singular point (green asterisk) in the sonic line $(1-U)^2+C^2=0$ (orange). The red curve (b) is the solution behind the reflected shock, and the green curve (v) is given by applying the shock jump conditions (\ref{eq:jump_conds}) on the blue curve.
	The dashed lines are not part of the physical solution. \textbf{(a,b)} $\omega=1$ and $\omega=1.85$, a reflected shock exists since there is an intersection between (b) and (v), as both $\omega$ are less than $\omega_r\approx1.88$ \textbf{(c)} $\omega=2.1>\omega_r$, curves (b) and (v) no longer intersect, so there is no reflected shock and the post-convergence flow is shock-free. \textbf{(d)} $\omega=3.8>\omega_b$, the converging shock takes infinite time to reach the origin.}
	\label{fig:reflected_UC}
\end{figure*}

From here on, the curve of the solution that was obtained by integrating from $x=0$ will be denoted (a), and the curve obtained by integrating from a small $y_0$ will be denoted (b). The two curves lie on opposite sides of the second sonic line $U-C=1$. The solution jumps from (a) to (b) by means of a shock. A third, virtual curve (v) is calculated by applying the Hugoniot jump conditions on (a):
\begin{equation}
\label{eq:jump_conds}
\begin{gathered}
1-U_v=\frac{\gm-1}{\gm+1}(1-U_a)+\frac{2C_a^2}{(\gm+1)(1-U_a)} \\
C_v^2=C_a^2+\frac{\gm-1}{2}\left((1-U_a)^2-(1-U_v)^2\right).
\end{gathered}
\end{equation}
The point where curve (v) crosses curve (b) corresponds to the shock point on (a). This determines the value of $x$ at the shock $B$, which in turn determines $k$ and allows the translation of the (b) curve from depending on $y$ to $x$. Thus a full description of the flow, including the reflected shock and the doubly-shocked flow, is obtained. Panels I and II of Figure~\ref{fig:reflected_UC} demonstrate the intersection of (b) and (v) in $U-C$ space.

It is interesting that for some values of $\omg$, there is no intersection between (b) and (v), so a reflected shock cannot exist at all. This is obviously the case when convergence takes infinite time ($\omg>\omg_b$ and $\lmb<0$), but the disappearance of the reflected shock also occurs for lower values of $\omg$, when $\lmb$ is still positive. The reflected shock disappears when there is a stagnation point in the $x<0$ flow solution, i.e. a radius outside of which the flow velocity is directed outwards even during the shock convergence phase. The lowest value of $\omg$ for which there is no reflected shock wave is denoted by $\omg_r$. Figure~\ref{fig:reflected_UC} (III and IV) shows examples of $U-C$ diagrams without a reflected shock, for $\omg_r<\omg<\omg_b$ and for $\omg>\omg_b$. 

\subsection{The gap}

For diverging spherical shocks $n=2,s=1$, Waxman and Shvarts \citep{waxman1993} found that there is a special region $3<\omg<\omg_g$ where type I solutions are invalid, and type II solutions that pass through the singular sonic point do not exist. Gruzinov \citep{gruzinov2003} found the correct solutions in this region and referred to them as type III solutions. Kushnir and Waxman \citep{kushnir_waxman2010} argued that these solutions are still type II solutions, but of a different variety. To emphasize the different properties of these solutions in contrast to ``regular" type II solutions, we will refer to them as ``gap solutions".

Kushnir and Katz \citep{kushnir_katz2014} studied the flow created by a piston in power-law density, which has the same solutions as the explosion problem when the shock is not decelerating $\lmb\le1$, and showed that a gap also exists for cylindrical geometry, but not for planar geometry. In this subsection, we study the gap from the point of view of the explosion problem, which provides more insight as to why a gap does not exist in planar flow.

As proven in appendix A, type I self-similar solutions only appear in the explosion problem with $\omg\le n+1$, so the transition between type I and type II solutions is only relevant when $s=1$. In planar geometry, this transition is smooth. $n=0$ implies that $\omg_h=1$ regardless of the value of $\gm$, and that is precisely the point where type I solutions' validity ends. Thus, there is no hollow region in any of the planar type I solutions. In addition, from (\ref{eq:sonic_point}) it can be seen that for $\omg=1,\lmb=1$ the sonic point is at $(U,C)=(1,0)$, the point where type I solutions with $\omg>\omg_h$ terminate. From these considerations, it can be understood that in planar explosions, type I solutions can transition into type II solutions continuously.

For $n>0$ the transition is not so simple. Since $n+1>\omg_h$, type I solutions at their validity limit $\omg=n+1$ always have a finite hollow region in the center. Type II solutions pass through the sonic point and terminate on $(U,C)=(\lmb,0)$, which cannot be a vacuum interface when $\lmb\ne1$. If the transition from type I to type II were immediate, then the hollow region would be closed abruptly when $\omg>n+1$. This contradicts the likely assumption that the physical solution changes continuously as a function of $\omg$. This is also in agreement with Waxman and Shvarts \citep{waxman1993}, who found numerically that a solution that passes through a singular point does not exist for values of $\omega$ slightly above $3$ (in spherical geometry).

\begin{figure}
	\centering
	\includegraphics[width=\columnwidth]{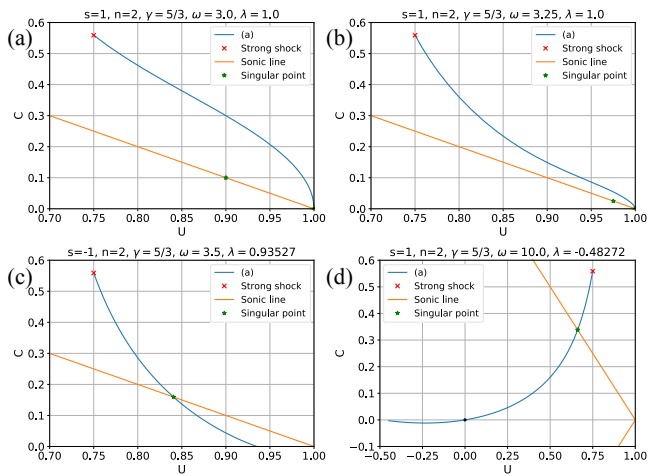}
	\caption{$U-C$ diagrams (same as in figure~\ref{fig:reflected_UC}) of solutions  to equation (\ref{eq:main_ODE}) for spherical diverging shocks with $\gamma=5/3$ and various $\omega$. \textbf{(a)} $\omega=3$, hollow type I solution. \textbf{(b)} $\omega=3.25<\omega_g$, a gap solution. \textbf{(c)} $\omega=3.5>\omega_g$, type II solution. \textbf{(d)} $\omega=10>\omega_c$, type II solution with finite-time shock divergence.}
	\label{fig:explosion_UC}
\end{figure}

The solution to this ostensible paradox was proposed by Gruzinov \cite{gruzinov2003} and further developed by Kushnir and Waxman \cite{kushnir_waxman2010}. Instead of type I transitioning immediately to a regular type II solution, there exists a ``gap" in which the solution is neither. In the gap $n+1<\omg<\omg_g$, the similarity exponent is independent of $\omega$ and is always $\lmb=1$. As $\omg$ approaches $\omg_g$, the hollow region shrinks until it vanishes at $\omg_g$ and the solution becomes a regular type II solution. Figure~\ref{fig:explosion_UC} shows the transition from type I to type II solutions in $U-C$ diagrams.

\section{Discussion of solutions and numerical confirmation}

%Section \ref{sec:self_similar_solution} described how to obtain self-similar solutions to the generalized shock problem in any of the various regions. Solutions were shown in $U-C$ space since it is the most useful approach for obtaining them, but it should be reminded that a full solution is given in terms of the similarity variable $x$, and therefore the profiles $U(x)$, $C(x)$ and $G(x)$ are of interest.

\begin{figure}
	\centering
	\includegraphics[width=\columnwidth]{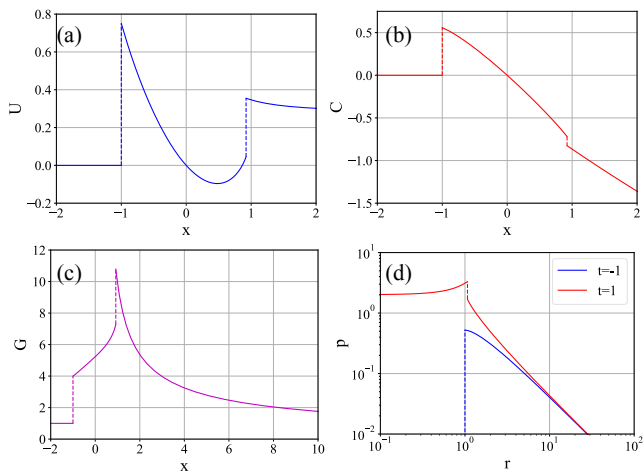}
	\caption{Throughout this figure, $s=-1$, $n=2$, $\gamma=5/3$, $\omega=1$. \textbf{(a)} $U(x)$. \textbf{(b)} $C(x)$. \textbf{(c)} $G(x)$. \textbf{(d)} The pressure as a function of radius, before (blue) and after (red) shock convergence, in logarithmic scale.}
	\label{fig:conv_w1_profiles}
\end{figure}

While in Section \ref{sec:self_similar_solution} we focused on the properties of the solution in the $U-C$ plane, here, we inspect the self-similar profile of the solutions as function of $x$.
Figure~\ref{fig:conv_w1_profiles} demonstrates %how these profiles look in
the case of a converging shock in spherical symmetry and $\gamma=5/3$, $\omega=1$. In addition to the $x$-profiles, panel IV shows the spatial profile of the pressure at two snapshots, before and after shock convergence. It can be seen how for very large radii, the two profiles approach one another; that is because the time difference between the snapshots is negligible, compared to the dynamical timescale at large radii.

\begin{figure}
	\centering
	\includegraphics[width=\columnwidth]{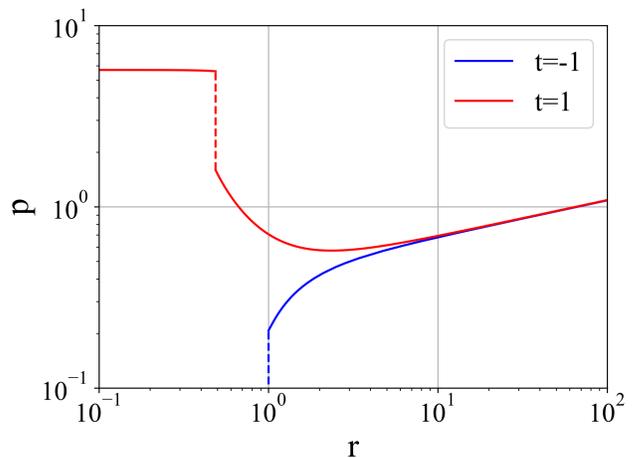}
	\caption{The pressure as a function of radius for $s=-1$, $n=2$, $\gamma=5/3$, $\omega=-2$, both before (blue) and after (red) shock convergence.}
	\label{fig:pr_conv_wm2}
\end{figure}

An interesting behavioral change that is not evident in $U-C$ diagrams can be observed by examining the radial pressure profile. When $\omg$ is very negative, the density around the center is very low, and thus even at the moment of shock convergence $t=0$, the pressure at the origin is zero. Despite this, a reflected shock exists for $t>0$. This is especially interesting since infinite pressure at the origin is usually taken as the justification for the existence of the reflected shock wave \citep{lazarus1981}. A condition for the pressure to vanish at the origin can be derived from the self-similar ansatz (\ref{eq:ansatz}). In the limit $t\to0$ ($x\to0$),
\begin{equation}
	P\sim\rho c^2 \sim r^{-\omg+2(1-\lmb)}.
\end{equation}
Therefore, when $\omg+2(\lmb-1)<0$, the pressure at the origin vanishes at $t=0$. The radial pressure profile of such a solution is shown in figure~\ref{fig:pr_conv_wm2}.

Numerical validation of all types of solutions in explosions has been published in literature \citep{waxman1993,waxman2010,kushnir_waxman2010}. Implosions and reflected shocks with $\omg\ne0$ were studied numerically in detail only recently by Giron et al. \citep{giron2021A,giron2021B}, who present them as a reference for validation and verification of numerical simulations of one dimensional hydrodynamic flow. However, only intermediate $\omg$ cases have been studied, i.e. where the pressure at shock convergence is infinite and a reflected shock exists.

Here, we demonstrate the convergence of the flow to the solutions in the various regions of the implosion problem. The simulation used in this work utilizes a Lagrangian scheme with quadratic artificial viscosity \citep{richtmyer_morton1957}, with additional linear artificial viscosity to smooth out small oscillations.

Since $n$ and $\gm$ don't change the qualitative features of the behavioral phases, all simulations were performed in spherical geometry $n=2$ and $\gm=\frac{5}{3}$. The solutions are tested in 4 different cases, each representing a different phase of behavior.
 
For $n=2$ and $\gm=\frac{5}{3}$, the value of $\omg$ below which the pressure in the origin vanishes during shock convergence is approximately -1.646. The next major behavioral changes are the disappearance of the reflected shock wave at $\omg_r\approx1.88$ and the end of finite-time shock convergence at $\omg_b\approx3.544$. Thus, the four cases that were chosen to be simulated are $\omg=\left\{ -2,1,3,5 \right\}$.

\begin{figure}
	\centering
	\includegraphics[width=\columnwidth]{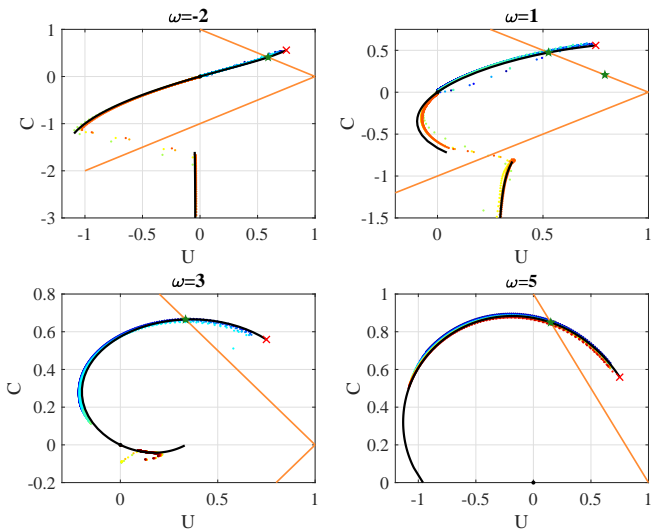}
	\caption{$U-C$ diagrams for $s=-1,n=2,\gm=\frac{5}{3}$ and $\omg\;\in\,\left\{-2,1,3,5\right\}$. The black line is the self-similar solution and the colored dots are taken from a simulation (same color dots are from the same time in the simulation). The orange lines are the sonic lines $C^2=(1-U)^2$, the green stars are the singular sonic points, and the red cross is the strong shock point.}
	\label{fig:sim_UC}
\end{figure}

Figure~\ref{fig:sim_UC} shows $U-C$ diagrams of the self-similar solution, as well as points from various times during the simulation. The $U,C$ values of the simulation are calculated according to equation (\ref{eq:ansatz}). It can be seen that in all cases, the self-similar solution agrees with the results of the simulation. The results shown here establish that the self-similar solutions describe the asymptotic behavior of the simulated flow well, and that all expected phase changes occur in the simulations.

\section{Approximate analytic results}

A general closed-form expression for the similarity exponent $\lmb(s,n,\gm,\omg)$ does not exist. Several approximations exist for converging shocks in uniform density \citep{stanyukovich1960,chisnell1957} $\lmb(s=-1,n,\gm,\omg=0)$. An approximation for planar shocks in variable density $\lmb(s=-1,n=0,\gm,\omg)$ is known \citep{sakurai1960}, but it is not valid for large values of $\omg$. In the case of diverging shocks, no attempts to approximate the similarity exponent in the type II flow region have been made. % It should be noted that Vishwakarma \citep{vishwakarma2005} shows a method for obtaining an approximation for the similarity exponent in converging shocks with arbitrary $\omg$, but this method requires solving a transcendental equation, does not give a closed-form expression, and does not work well for large $\omg$.

Nonetheless, it is possible to derive complementary approximations, which together add up to a general approximate expression for $\lmb(s,n,\gm,\omg)$.

\subsection{Approximate similarity exponent for converging shocks}

Appendix B shows the derivation of an approximate expression for $\lmb(s=-1,n,\gm,\omg)$ that generalizes previous analytic estimates \citep{chisnell1957,sakurai1960} using ideas from Whitham's geometrical shock theory \citep{whitham_waves}. This approximation works well around $\omg=0$ and for negative values of $\omg$, but breaks down for higher values. The reason for this is that the approximation assumes a uniform flow behind the converging shock, which is far from being the case when $\omg$ reaches $\omg_r$, and a stagnation point exists in the flow.

\begin{figure}
    \centering
    \includegraphics[width=\columnwidth]{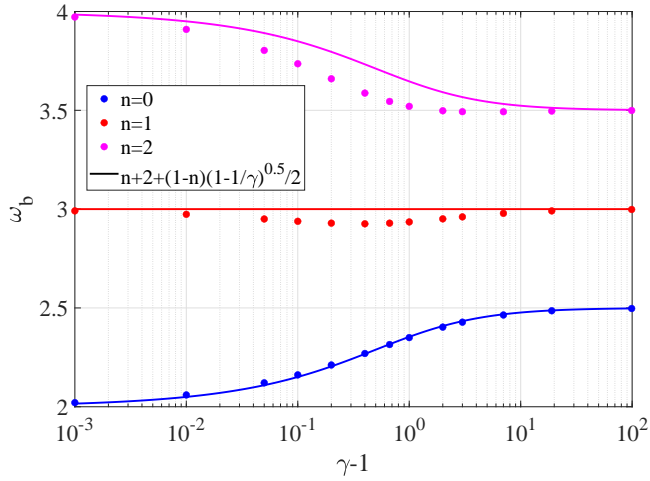}
    \caption{Plots of $\omg_b$ as functions of $\gm$ for various $n$. The points are numerical results, and the lines are the analytical estimate in equation (\ref{eq:wb_anal}).}
    \label{fig:wb_vs_gm}
\end{figure}

\begin{figure}
    \centering
    \includegraphics[width=\columnwidth]{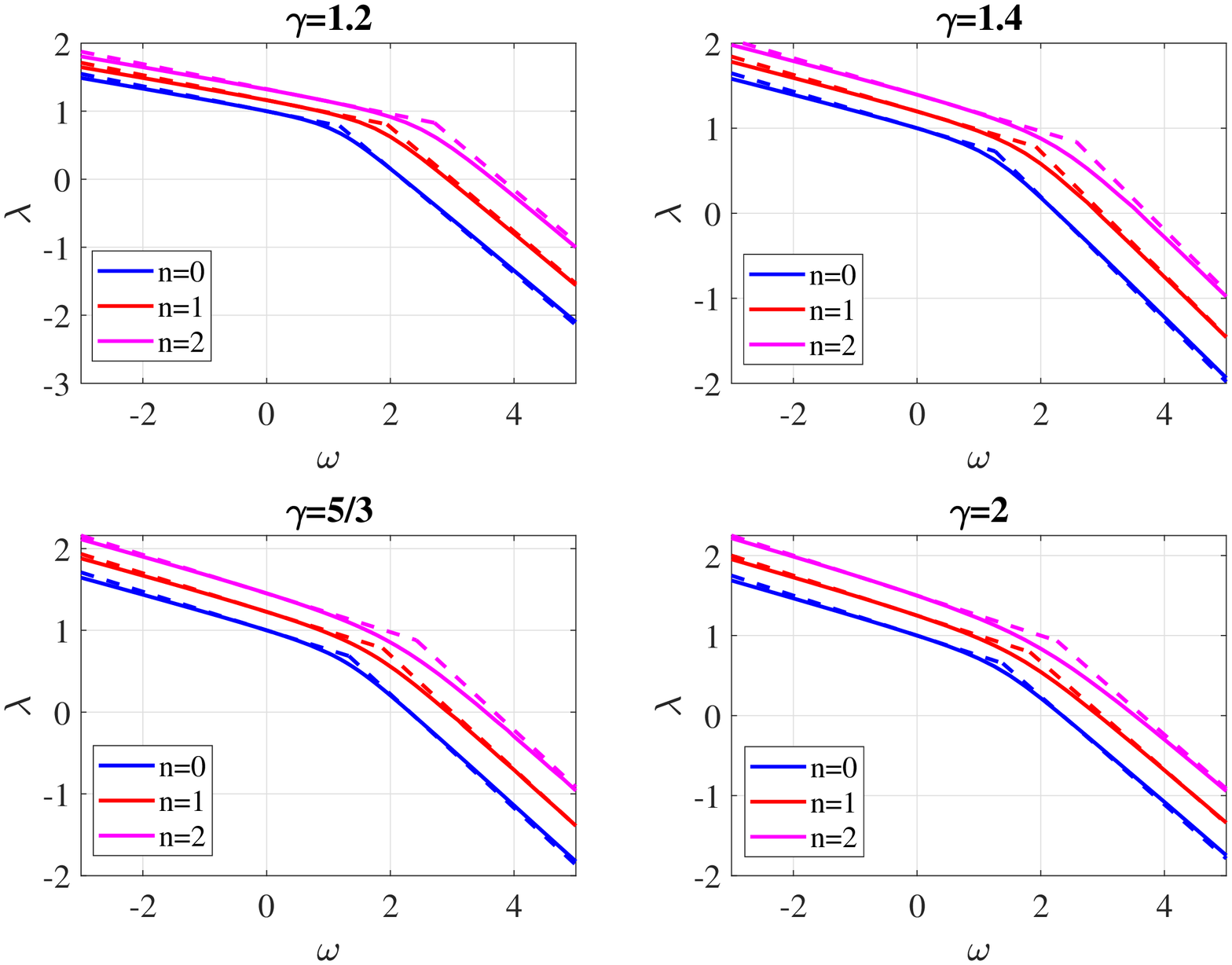}
	\caption{Plots of $\lmb(\omg)$ as a function of $\omg$ for various $\gm$ and $n$, in converging shocks $s=-1$. The solid lines are numerical results, and the dashed lines are the analytic estimate in equation (\ref{eq:lmb_conv_approx}).}
	\label{fig:conv_lmb_vs_omg}
\end{figure}

Numerical results show that $\lmb(s=-1,n,\gm,\omg)$ as a function of $\omg$ is fairly similar to two straight lines with different slopes $\eta_{1,2}=-\frac{d\lmb}{d\omg}$ that meet around $\omg_r$. The slope and the offset of the line coming from $\omg=-\infty$ are given approximately by the equation (\ref{eq:whitham_anal_lmb}) from appendix B:
\begin{equation}
\label{eq:whitham_in_text}
\begin{gathered}
	\eta_1=\left( 2+\sqrt{\frac{2\gm}{\gm-1}} \right) ^{-1} \\
	\lmb(\omg=0)=1+\frac{n}{1+\frac{2}{\gm}+\sqrt{\frac{2\gm}{\gm-1}}}.
\end{gathered}
\end{equation}
Equation (\ref{eq:eta2_approx}) from appendix C gives an estimate for the slope of the second line
\begin{equation}
\label{eq:eta2_in_text}
	\eta_2=1-0.4\left( \frac{\gm-1}{\gm} \right) ^{0.3}.
\end{equation}
We now obatin an estimate for $\omg_b$, where $\lmb(\omg_b)=0$, in order to find the switch point between the two lines. The numerical results in figure~\ref{fig:wb_vs_gm} show that $\omg_b\to n+2$ when $\gm\to1$ and $\omg_b\to \frac{n+5}{2}$ when $\gm\to\infty$. A simple estimate that conforms to these limits is
\begin{equation}
\label{eq:wb_anal}
	\omg_b=n+2+\frac{1-n}{2}\sqrt{\frac{\gm-1}{\gm}}.
\end{equation}
This estimate works exceptionally well for $n=0$. For $n=1$ it is accurate up to 3\% and for $n=2$ up to 4\%. Equations (\ref{eq:whitham_in_text}), (\ref{eq:eta2_in_text}) and (\ref{eq:wb_anal}) can be combined to derive an approximation for $\lmb$ in converging shocks:
\begin{equation}
\label{eq:lmb_conv_approx}
\begin{gathered}
	\lmb(s=-1,n,\gm,\omg)=
	\begin{cases}
	\lmb(0)-\eta_1\omg & \omg<\omg_s\\
	\eta_2(\omg_b-\omg) & \omg>\omg_s
	\end{cases}, \\
	\omg_s=\frac{\eta_2\omg_b-\lmb(0)}{\eta_2-\eta_1}.
\end{gathered}
\end{equation}
Figure~\ref{fig:conv_lmb_vs_omg} shows a comparison of this estimate to numerical results, which are in agreement with previous works.\citep{sakurai1960,sharma1995,toque2001,madhumita2003,giron2021A}

\subsection{Approximate similarity exponent for diverging shocks}
\label{sec:div_lmb}

Unlike converging shocks, diverging shocks have analytic type I solutions for $\omega<n+1$ where $\lmb=\frac{n+3-\omg}{2}$ is an exact relation. In ``the gap" $n+1<\omg<\omg_g$, $\lmb=1$ is known without the need for calculation. For $\omg>\omg_g$, the solution is type II and there is no exact analytic expression for $\lmb$. Several results can be utilized in order to get an approximate expression for $\lmb$ in this region.

First, there is an interesting connection between converging shocks with very large negative $\omg$ to diverging shocks with very large positive $\omg$. For both these cases, the density drops so sharply that its importance to the solution overshadows geometrical effects. In other words, when $s\omg$ is very large, the solutions depend weakly on $s$ and $n$ before shock convergence/divergence. That is because the length scale for changes in the flow variables behind the shock is $\frac{R}{\omg} \ll R$. As implied by this, the slope $\frac{d\lmb}{d\omg}$ is the same for diverging shocks as it is for converging shocks, and the approximation in Appendix B is applicable.
\[
	\lim_{\omg\to\infty}\frac{d\lmb}{d\omg}=-\left(2+\sqrt{\frac{2\gm}{\gm-1}}\right)^{-1}=-\eta_1(\gm).
\]
Numerical results show that the slope of $\lmb(\omg)$ does not change significantly after $\omg_g$, so for $\omg>\omg_g$, $\lmb=\eta_1(\omg_c-\omg)$ is a good approximation. In order to use this, estimates for $\omg_g$ and $\omg_c$ need to be introduced.

\begin{figure}
    \centering
    \includegraphics[width=\columnwidth]{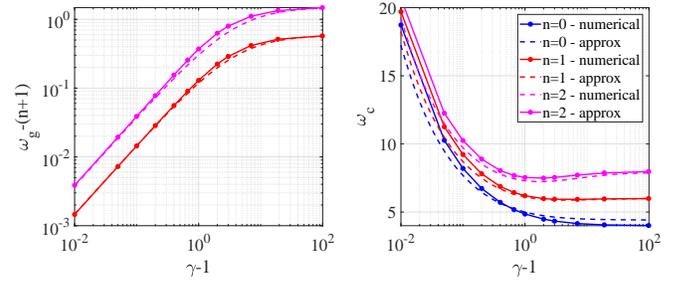}
    \caption{Plots of $\omg_g$ (left) and $\omg_c$ (right) as functions of $\gm$ for various $n$. The points are numerical results, and the dashed lines are the analytical estimates in equations (\ref{eq:wg_anal}) and (\ref{eq:wc_anal}).}
    \label{fig:wg_wc_vs_gm}
\end{figure}

\begin{figure}
    \centering
    \includegraphics[width=\columnwidth]{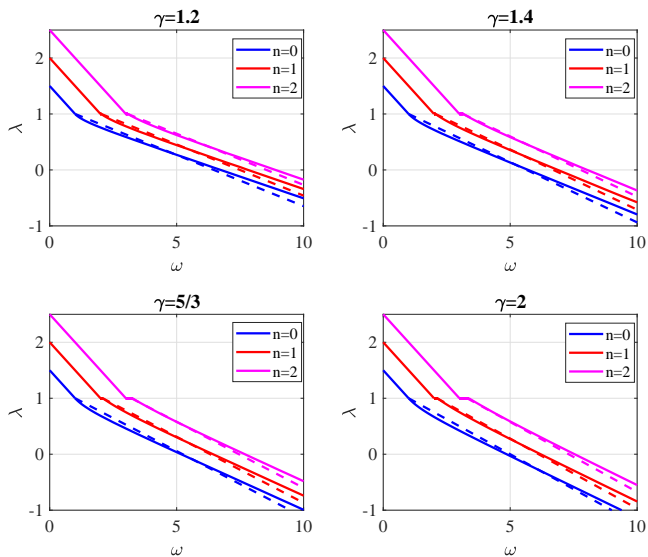}
    \caption{Plots of $\lmb(\omg)$ as functions of $\omg$ for various $\gm$ and $n$, in diverging shocks $s=1$. The solid lines are numerical results, and the dashed lines are the analytical estimate in equation (\ref{eq:lmb_exp_anal}).}
    \label{fig:div_lmb_vs_omg}
\end{figure}

An upper bound for $\omg_g$ and a lower bound for $\omg_c$ can be obtained by considering that for a particular choice of $\omg$, there exists an analytic type II solution. For $\omg=n+1+(n+3)\frac{\gm-1}{\gm+1}$, the exact value of $\lmb$ is $\frac{2}{\gm+1}$. Since this value of $\lmb$ is between 0 and 1, it proves the inequality
\begin{equation}
	0\le\omg_g-(n+1)<(n+3)\frac{\gm-1}{\gm+1}<\omg_c-(n+1).
\end{equation}
This bound becomes tighter for $\omg_g$ in low $\gm$ and tighter for $\omg_c$ in high $\gm$. This bound proves that the gap width tends to zero as $\gm\to1$ and that $\omg_c\to 2n+4$ as $\gm\to\infty$. Figure~\ref{fig:wg_wc_vs_gm} shows numerical results for $\omg_g$ and $\omg_c$. It can be seen that the simple analytic expression
\begin{equation}
\label{eq:wg_anal}
\begin{gathered}
	\omg_g-(n+1)\approx\alpha_n\frac{\gm-1}{\gm+3}, \\
	\alpha_n=
	\begin{cases}
	0 & n=0\\
	0.5905 & n=1\\
	1.5148 & n=2
	\end{cases}.
\end{gathered}
\end{equation}
provides a decent approximation for all values of $\gm$ and $n$. It is especially accurate at the limits $\gm\to1$ and $\gm\to\infty$. Combining this with the assumption that the slop $\frac{d\lmb}{d\omg}=-\eta_1$ is constant, we get the following approximation for $\omg_c$
\begin{equation}
\label{eq:wc_anal}
	\omg_c\approx\omg_g+\eta_1^{-1}\approx n+3+\alpha_n\frac{\gm-1}{\gm+3}+\sqrt{\frac{2\gm}{\gm-1}}.
\end{equation}
which is also shown in Figure~\ref{fig:wg_wc_vs_gm}. Combining all results, an estimate for $\lmb(s=1,n,\gm,\omg)$ is obtained:
\begin{equation}
\label{eq:lmb_exp_anal}
\begin{gathered}
	\lmb=
	\begin{cases}
	1-\frac{\omg-(n+1)}{2} & \omg-(n+1)\le 0 \\
	1 & 0<\omg-(n+1)\le \Delta\omg_g \\
	1-\frac{\omg-(n+1)-\Delta\omg_g}{2+\sqrt{\frac{2\gm}{\gm-1}}} & \Delta\omg_g<\omg-(n+1)
	\end{cases}\\
	\Delta\omg_g(n,\gm)=\alpha_n\frac{\gm-1}{\gm+3}.
\end{gathered}
\end{equation}
Figure~\ref{fig:div_lmb_vs_omg} shows a comparison of this estimate to numerical results.

\subsection{Connecting different geometries}

As discussed in sec. \ref{sec:div_lmb}, when $s\omg\to\infty$ the solution close to the shock is similar in all geometries. As a consequence of that, it was shown that $\eta_1=-\lim_{s\omg\to\infty} \frac{d\lmb}{d\omg}$ is independent of $s$ and $n$. This relation can be expanded using physical arguments, to show a deeper connection between all geometries.

The symmetry parameter $n$ can be eliminated from the equation of continuity (first equation in (\ref{eq:1dflow})) if instead of the volume density $\rho$, we use the linear density $g=\rho r^n$. While this transformation does not eliminate $n$ from all flow equations, it highlights a major effect of the symmetry $n$. Using this idea, we argue that instead of $\omg=-\frac{d\log\rho}{d\log r}$, the parameter  $\frac{d\log g}{d\log r}=n-\omg$ better governs the flow behavior. To combine diverging and converging shocks, it should be noted that the logarithmic derivative needs to be taken in the direction of shock propagation
\begin{equation}
     s\frac{d\log g}{d\log r}=s(n-\omg).
\end{equation}

Taking this line of thinking further, we consider the rate at which the shock propagates in terms of mass rather than radius:
\[
    \frac{dM}{dt}=\frac{dM}{dR}\frac{dR}{dt} \sim R^{n-\omega}R^{1-\lmb}.
\]
Instead of $\lmb=1-\frac{d\log \dot{R}}{d\log R}$, a similarity exponent, signifying the mass propagation rate's dependence on radius, would be
\begin{equation}
    s\frac{d\log \dot{M}}{d\log R} = s(n+1-\omg-\lmb).
\end{equation}

\begin{figure}
    \centering
    \includegraphics[width=\columnwidth]{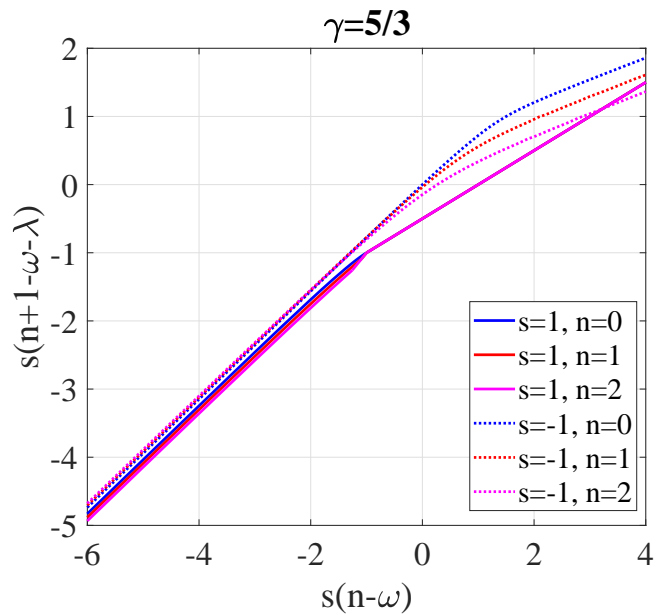}
    \caption{$s(n+1-\omega-\lambda)$ as a function of $s(n-\omega)$ in all geometries $s$ and $n$ for $\gamma=5/3$.}
    \label{fig:generalized_lmb_omg}
\end{figure}

Figure~\ref{fig:generalized_lmb_omg} shows $s(n+1-\omg-\lmb)$ as a function of $s(n-\omg)$ in all geometries $s$ and $n$. For large negative values of $s(n-\omg)$, the behavior is very similar in all geometries. In these cases, the shock is accelerating and thus interacts weakly with the flow behind it. This makes the steep linear density $g\sim r^{n-\omg}$ gradient the dominant influence on the flow, supporting the physical arguments in this section.

On the other hand, the behavior varies substantially for positive values of $s(n-\omg)$, because the shock is decelerating and a larger portion of the flow needs to be taken into account to determine $\lmb$. 
For diverging shocks, $s=1$, in all symmetries $n$, the lines in figure~\ref{fig:generalized_lmb_omg} coincide perfectly, for $n-\omg>-1$, the type I region. This is because the combinations we chose $(n-\omega)$ and $(n+1-\omega-\lambda)$ are compatible with conservation of energy.

\section{Conclusion}
This article has shown how the strong explosion and the strong implosion problems can be treated as two cases of the same generalized problem, of a self-similar strong shock propagating in an ideal gas with initial power-law density profile $\rho\sim r^{-\omg}$. The generalized problem is defined by two logical parameters ($s$ and $n$) and two real parameters ($\gm$ and $\omg$). $s$ and $n$ define the geometry - the shock's direction of propagation and its symmetry, $\omg$ defines the steepness of the density profile and $\gm$ defines the properties of the gas itself.

Parameter space defines three regions, according to the type of the self-similar solution. Type I solutions are found when $s=1$ (explosion) and $\omg<n+1$, type II solutions appear when $s=1$ and $\omg>\omg_g$ or whenever $s=-1$ (implosion). The intermediate region in explosions contains Gruzinov's solutions \citep{gruzinov2003}.

The behavior of solutions as a function $\omg$ can be understood by recognizing the special values of $\omg$ at which the nature of the solutions changes qualitatively. In the explosion case, these values are
\begin{itemize}
	\item $\omg_h=1+\frac{3-\gm}{\gm+1}n$, where a hollow region appears in the center.
	\item $n+1$, where the outgoing shock is no longer decelerating, but instead has constant speed.
	\item $\omg_g$, where the hollow region in the center is closed again, and the shock becomes accelerating.
	\item $\omg_c$, where the shock reaches infinity in finite time, after which the flow is shock-free.
\end{itemize}
In the implosion case, these values are
\begin{itemize}
	\item $\omg_r$, where the in-going shock no longer creates a reflected shock after its convergence. This is also where a stagnation point appears in the converging flow.
	\item $\omg_b$, where the converging shock takes infinite time to reach the center.
\end{itemize}
It was also found that an additional, more minor qualitative change happens for low enough $\omg$, where the pressure at $r=0$ vanishes at shock convergence instead of being infinite.

This article has shown estimates for $\omg_g,\omg_c$ and $\omg_b$ for the first time in literature, as well as an approximate closed expression for a full description of $\lmb(s,n,\gm,\omg)$. While the estimates hold well for the entirety of parameter space, some of them are phenomenological and thus might not be applicable in further generalizations or variations of the strong shock problem.

The connection between the diverging and converging shock problems was first demonstrated by Waxman and Shvarts \citep{waxman1993} by using ideas from Guderley's converging shock solution \citep{guderley1942}. The two problems can be described by the same set of self-similar ODEs, and the solution is found by choosing $\lmb$ such that it passes through a singular sonic point. In this paper the connection is further established by showing that the converging and diverging problems can be treated as two cases of a generalized problem with the parameter $s$.
A deeper physical connection between different geometries was established by considering the rate at which shocks accumulate mass instead of their speed. This connection is especially meaningful for large values of $s(\omg-n)$, where all geometries have the same flow solution near the shock front.

\section*{Acknowledgement}
This research was partially supported by an ISF grant. We thank I. Giron and M. Krief for sharing with us the results of their current work, and for providing us with the method for deriving the reflected shock solution in sec. \ref{sec:reflected_shocks}.

%\newpage
\appendix

\section*{Appendix A - Proof that type I self-similar flow is impossible for converging shocks}
\renewcommand\theequation{A\arabic{equation}}
\setcounter{equation}{0}

This appendix provides a proof that there is no combination of $n,\gamma,\omega$ such that the appropriate self-similar solution for a converging shock is of the first kind. The proof is divided into two parts; first, proof that bounded type I solutions are impossible; second, proof that unbounded type I solutions are impossible.

Let us assume that the solution in bounded, i.e. the entirety of the fluid in the solution is contained in a finite volume that scales with the shock front location. The fluid-vacuum interface must have $(U,C)=(1,0)$, for the vacuum interface to stay at a constant $x_b$ (this is required for the solution to be self-similar) and for the fluid to have zero pressure at the interface.

The solution starts at the strong shock point $(U,C)=\left(\frac{2}{\gamma+1},\frac{\sqrt{2\gamma(\gamma-1)}}{\gamma+1}\right)$ and needs to advance towards the vacuum interface point $(U,C)=(1,0)$. Regardless of whether $\lambda>0$ (the shock is converging in finite time and $0>x_b>-1$) or $\lambda<0$ (the shock is converging in infinite time and $x_b>1$), this requires that $dU/dx>0$ between the shock point and the vacuum interface point. In both cases, $\lambda x<0$ and $\Delta<0$ at the shock point. Therefore, from equation (\ref{eq:main_ODE}), the sign of $\Delta_1$ at the shock point must be positive.
\small
\begin{equation*}
\begin{gathered}
	\Delta_1=\left[\frac{\omega+2(\lambda-1)}{\gamma}-(n+1)U\right]C^2+U(1-U)(\lambda-U)=\\
	%\left[\frac{\omega+2(\lambda-1)}{\gamma}-2\frac{n+1}{\gamma+1}\right]\frac{2\gamma(\gamma-1)}{(\gamma+1)^2}+\frac{2}{\gamma+1}\frac{\gamma-1}{\gamma+1}(\lambda-\frac{2}{\gamma+1})=\\
	\frac{2(\gamma-1)}{(\gamma+1)^3}\left[(\gamma+1)(\omega+3\lambda-2)-2((n+1)\gamma+1)\right]>0.
\end{gathered}
\end{equation*}
\normalsize
When the solution is type I, energy is conserved so $\lambda=\frac{n+3-\omega}{2}$. Plugging this into the inequality gives
\begin{equation*}
	\omega<1+n\frac{3-\gamma}{\gamma+1}.
\end{equation*}
It can be seen from this that $\omega$ cannot be greater than $n+1$, since it would require $\gamma<1$ which is not physical. Hence, $\omega<n+1$ which means that the mass of all matter outside a sphere around the origin is infinite:
\begin{equation}
\begin{gathered}
	M(r>R)\sim\int_R^\infty\rho r^n dr \\
	\sim\int_1^\infty\xi^{n-\omega} d\xi > \int_1^\infty \frac{d\xi}{\xi} = \infty.
\end{gathered}
\end{equation}
Recall that the solution is bounded, therefore an infinite mass has a non-zero velocity. The similarity exponent is $\lambda=\frac{n+3-\omega}{2}>1$, so the shock is accelerating towards the origin. Because of the self-similarity, the vacuum interface where the mass is infinite is also accelerating, which is impossible since no force acts on it from the side of the vacuum. This argument is similar to the argument used by Gruzinov \citep{gruzinov2003} to prove that type I solutions are not valid for explosions with $\omega>n+1$. This concludes the proof that bounded type I solutions are impossible for all geometries and all values of $\omega$.

Now let us assume that the flow is unbounded, i.e. there is no vacuum interface. The arguments for the explosion analytic type I solution are also valid for an implosion problem. Thus, we can use the relation between $C$ and $U$ (\ref{eq:type1_CU})
\begin{equation}
	C^2=\frac{\gamma(\gamma-1)(1-U)U^2}{2(\gamma U-1)}.
\end{equation}
This curve does not pass through the origin $U=C=0$, and therefore $t=0$ cannot be included in the solution. This means that the solution must describe infinite time convergence, and therefore $\lmb<0$ and $\omg>n+3$. There exists a value $\omg_r$ after which there is no shock reflection even when $\lmb>0$ (in a type II solution). When $\omg>\omg_r$, there is a stagnation point ($U=0$) in the flow. In type I solutions, $U>\frac{1}{\gamma}$ so there cannot be a stagnation point. Had type I solutions been valid for some $\omg>n+3>\omg_r$, there would have been a stagnation point since the density increases even more steeply (the situation is closer to a shock hitting a wall). This leads to the conclusion that type I solutions are never valid in converging shocks.

%\newpage
\section*{Appendix B - Geometrical shock theory}
\renewcommand\theequation{B\arabic{equation}}
\setcounter{equation}{0}

This appendix shows how to derive analytical approximations for $\lambda(n,\omega)$ for converging shocks using Whitham's ``Geometrical Shock Theory". The derivation is based on chapter 8 in \cite{whitham_waves}, combining the methods for shock propagation down a nonuniform tube and for shock propagation through a stratified layer.

We consider flow in a tube with varying cross-section area $A(r)$ and varying initial density $\rho_0(r)$. Assuming that the flow is one-dimensional (which is a good approximation if $A(r)$ varies slowly enough), the flow equations can be written as
\begin{equation}
\label{eq:whitham_1dflow}
    \begin{gathered}
        \rho_t+u\rho_r+\rho u_r+\rho u \frac{A'}{A}=0,\\
        u_t+uu_r+\frac{p_r}{\rho}=0,\\
        p_t+up_r-c^2(\rho_t+u\rho_r)=0.
    \end{gathered}
\end{equation}
This set of equation is equivalent to the regular one-dimensional flow equations (\ref{eq:1dflow}) when $A(r)\propto r^n$, since $A'/A=n/r$. Let us assume that a shock wave propagates through this tube, and that the fluid is initially at rest. The unperturbed flow variables will be denoted with a subscript 0, and flow variables immediately after the shock front will be denoted with a subscript 1. Hugoniot's jump conditions give
\begin{equation}
    \begin{gathered}
        u_1=c_0\frac{2}{\gamma+1}\left(M-\frac{1}{M}\right),\\
        p_1=\rho_0 c_0^2\left(\frac{2}{\gamma+1}M^2-\frac{\gamma-1}{\gamma(\gamma+1)}\right),\\
        \rho_1=\rho_0\frac{(\gamma+1)M^2}{(\gamma-1)M^2+2}.
    \end{gathered}
\end{equation}
Assuming that $A$ does not change much in relation to some initial value $A_0$, equations (\ref{eq:whitham_1dflow}) can be linearized to calculate perturbation from $\rho_1,p_1,u_1$.
\begin{equation}
\label{eq:linearized_whitham_1dflow}
    \begin{gathered}
        \rho_t+u_1\rho_r+\rho_1 u_r+\rho_1 u_1 \frac{A'}{A_0}=0,\\
        u_t+u_1 u_r+\frac{p_r}{\rho_1}=0,\\
        p_t+u_1 p_r-c_1^2(\rho_t+u_1\rho_r)=0.
    \end{gathered}
\end{equation}
This can be rewritten as characteristics equations:
\small
\begin{equation}
\label{eq:whitham_characteristics}
\begin{gathered}
    C_+: \left[\partial_t+(u_1+c_1)\partial_r\right](p+\rho_1 c_1 u)+\rho_1 c_1^2 u_1 \frac{A'}{A_0}=0,\\
    C_-: \left[\partial_t+(u_1-c_1)\partial_r\right](p-\rho_1 c_1 u)+\rho_1 c_1^2 u_1 \frac{A'}{A_0}=0,\\
    C_0: \left[\partial_t+u_1\partial_r\right](p-c_1^2\rho)=0.
\end{gathered}
\end{equation}
\normalsize
From now on, we will only be interested in the $C_+$ characteristic equation. To investigate how the flow variables immediately behind the shock change as the shock advances, the $C_+$ equation cannot be used as it is, since any shock is subsonic in relation to the fluid behind it, and thus the $C_+$ characteristic reaches over the shock front into the unperturbed fluid. Because of this, we will need to use a derivative along the shock instead of along $C_+$.
\begin{equation*}
    \partial_t+(u_1+c_1)\partial_r=\dot{R}\partial_R
\end{equation*}
Hence, the derivative of $A$ in relation to $R$ is given by
\begin{equation*}
    \frac{dA}{dr}=\frac{\dot{R}}{u_1+c_1}\frac{dA}{dR}
\end{equation*}
Plugging this into the $C_+$ equation in (\ref{eq:whitham_characteristics}) and dropping the subscripts gives
\begin{equation}
\label{eq:whitham_ode}
    \frac{dp}{dR}+\rho c \frac{du}{dR} + \frac{\rho c^2 u}{u+c}\frac{1}{A}\frac{dA}{dR}=0
\end{equation}
In the limit of a strong shock $M\to\infty$, Hugoniot's jump conditions amount to
\begin{equation}
\begin{gathered}
\label{eq:whitham_hugoniot}
    u=\frac{2}{\gamma+1}\dot{R},\quad \rho=\frac{\gamma+1}{\gamma-1}\rho_0,\\ p=\frac{2}{\gamma+1}\rho_0\dot{R}^2,\quad c^2=\frac{2\gamma(\gamma-1)}{\gamma+1}\dot{R}^2.
\end{gathered}
\end{equation}
Combining (\ref{eq:whitham_ode}) and (\ref{eq:whitham_hugoniot}) yields
\begin{equation}
    \frac{1}{\rho_0}\frac{d\rho_0}{dR}+\left(2+\sqrt{\frac{2\gamma}{\gamma-1}}\right)\frac{1}{\dot{R}}\frac{d\dot{R}}{dR}+\frac{1}{\frac{1}{\gamma}+\sqrt{\frac{\gamma-1}{2\gamma}}}\frac{1}{A}\frac{dA}{dR}=0.
\end{equation}
In this equation, all derivatives are known:
\begin{equation*}
    \frac{R}{\rho_0}\frac{d\rho_0}{dR}=-\omega,\quad \frac{R}{\dot{R}}\frac{d\dot{R}}{dR}=1-\lambda,\quad
    \frac{R}{A}\frac{dA}{dR}=n.
\end{equation*}
Finally, an analytical approximate value for $\lambda$ is obtained
\begin{equation}
\label{eq:whitham_anal_lmb}
    \lambda=1+\frac{n}{1+\frac{2}{\gamma}+\sqrt{\frac{2\gamma}{\gamma-1}}}-\frac{\omega}{2+\sqrt{\frac{2\gamma}{\gamma-1}}}.
\end{equation}

%\newpage
\section*{Appendix C - The limit $|\omega|\to\infty$}
\renewcommand\theequation{C\arabic{equation}}
\setcounter{equation}{0}

In the limit where $|\omg|\to\infty$, solutions tend to a certain curve in $U-C$ space. This happens since $\lmb$ becomes proportional to $\omg$ in such a way that eliminates $\omg$ from equation (\ref{eq:dudC}). $\lmb$ becomes linear in $\omg$, and can be written as $\lmb\to-\eta\omg$. This simplifies equation (\ref{eq:dudC}) a bit, resulting in:
\begin{equation}
\label{eq:dUdC_eta}
	\frac{dU}{dC}=\frac{1-U}{C}\frac{\frac{1-2\eta}{\gm}C^2-\eta U(1-U)}{\frac{2\eta+\gm-1}{2\gm}C^2-\eta\left(1+\frac{\gm-3}{2}U\right)(1-U)}.
\end{equation}

It is important to note that equation (\ref{eq:dUdC_eta}) is valid under the assumption that $U/\omg\ll1$, which is not generally true, but is applicable around the strong shock and the singular sonic point. It is also worth mentioning that equation (\ref{eq:dUdC_eta}) does not depend on $n$. This result makes sense, since when $\omg$ is very large, it is to be expected that near the shock, the solution would not notice the geometry, as the density gradient is much more dominant.

Utilizing the same argument, $s$ is also irrelevant by itself in this limit. It does not matter if the shock is in-going or out-going, it only matters if the shock is traveling with or against the density gradient. For this reason, diverging shocks with $\omega\to\infty$ have the same solution as converging shocks with $\omega\to-\infty$ before shock divergence/convergence. In other words, there is a single solution in the limit $s\omg\to\infty$. At $t>0$, the solutions are no longer the same since $U/\omg\ll1$ stops holding at some point.

The same cannot be said about the limit $s\omg\to-\infty$. In this case there is a fundamental difference between diverging and converging shocks - the first is type I and the latter is type II. In addition, while the approximation shown in appendix B works well in the limit $s\omg\to\infty$ and gives an estimate for $\eta_1=\lim_{s\omg\to\infty}\left(-\frac{\lmb}{\omg}\right)$, it is not valid in the limit $s\omg\to-\infty$, since the shock interacts strongly with the flow behind it.

\begin{figure}
	\centering
	\includegraphics[width=\columnwidth]{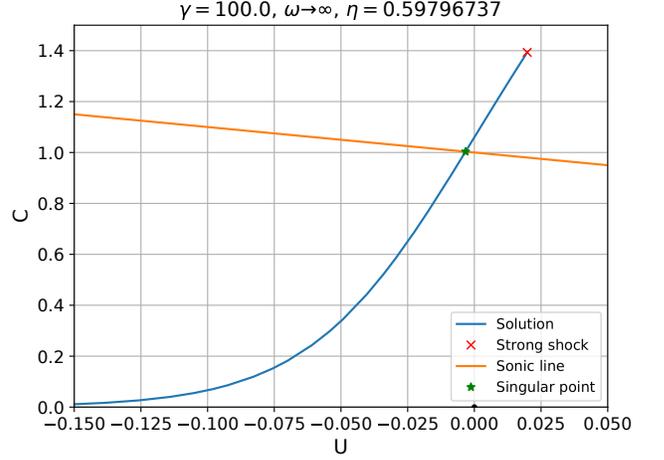}
	\caption{$U-C$ diagram of $\gm=100$ and $\omg\to\infty$ for a converging shock $s=-1$.}
	\label{fig:eta_g100}
\end{figure}

Diverging shocks with $\omg\to-\infty$ have a simple exact formula for $\lmb$ (\ref{eq:st_exponent}), but even finding an approximation for $\eta_2=\lim_{\omg\to\infty}\left(-\frac{\lmb}{\omg}\right)$ in converging shocks is very difficult. Some insight can be gained by recognizing that when $\gm$ is very large, the $U-C$ curve between the shock point and the singular sonic point is close to a straight line (see Figure~\ref{fig:eta_g100}).

Using equation (\ref{eq:sonic_point}) with $n=0$, the singular sonic point is at
\begin{equation}
\label{eq:eta_sonic_point}
	C_*=\frac{\eta\gm}{1+\eta(\gm-2)},\quad U_*=\frac{1-2\eta}{1+\eta(\gm-2)}.
\end{equation}
Assuming the solution curve is a straight line from the strong shock point \ref{eq:strong_shock_point} to the singular point (\ref{eq:eta_sonic_point}), equation (\ref{eq:dUdC_eta}) gives
\begin{equation}
	\frac{\frac{2}{\gm+1}-\frac{1-2\eta}{1+\eta(\gm-2)}}{\frac{\sqrt{2\gm(\gm-1)}}{\gm+1}-\frac{\eta\gm}{1+\eta(\gm-2)}}=\sqrt{\frac{\gm-1}{2\gm}} \frac{1-3\eta} {\frac{\gm-1}{2}-\eta(\gm-2)}.
\end{equation}
This amounts to a quadratic equation for $\eta$ which can be solved analytically. Its roots are
\begin{equation}
\label{eq:eta12_straight}
	%\eta_{1,2}=\frac{4(\gm-1)+\sqrt{2\gm(\gm-1)} \pm \sqrt{2(\gm-1)\left( 5\gm-2\sqrt{2\gm(\gm-1)} \right)}} {4(\gm-2)+6\sqrt{2\gm(\gm-1)}}.
	\eta_{1,2}=\frac{4+\sqrt{\frac{2\gm}{\gm-1}} \pm \sqrt{2\left( \frac{5\gm}{\gm-1}-2\sqrt{\frac{2\gm}{\gm-1}} \right)}} {4-\frac{4}{\gm-1}+6\sqrt{\frac{2\gm}{\gm-1}}}.
\end{equation}
At the limit where this approximation should be best, $\gm\gg1$, the result is
\begin{equation}
	\eta_{1,2}\to\frac{4+\sqrt2\pm\sqrt{2(5-2\sqrt2)}}{4+6\sqrt2} = \{0.2667,0.6006\}.
\end{equation}

\begin{figure}
	\centering
	\includegraphics[width=\columnwidth]{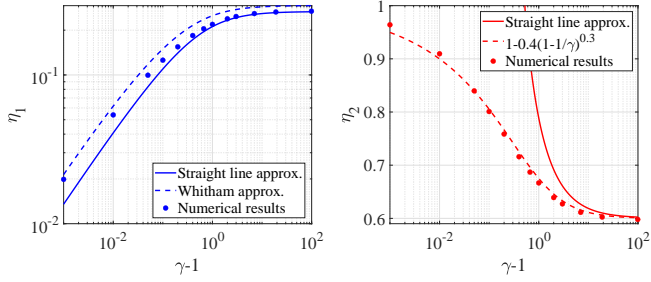}
	\caption{Plots of $\eta_1$ ($s\omg\to\infty$, left) and $\eta_2$ ($\omg\to\infty$ in $s=-1$, right) compared to analytic estimates. The straight line approximation refers to equation (\ref{eq:eta12_straight}) and Whitham's approximation refers to equation (\ref{eq:whitham_anal_lmb}).}
	\label{fig:eta_vs_gamma}
\end{figure}

Figure~\ref{fig:eta_vs_gamma} shows numerical results for $\eta_{1,2}$ in comparison to equation (\ref{eq:eta12_straight}) and to equation (\ref{eq:whitham_anal_lmb}) from Appendix B. It can be seen that both (\ref{eq:eta12_straight}) and (\ref{eq:whitham_anal_lmb}) provide reasonable estimates for $\eta_1$. The same cannot be said about $\eta_2$, where equation (\ref{eq:eta12_straight}) provides the limiting value for $\gm\to\infty$ but is otherwise a terrible approximation. A simple phenomenological formula is found to be a fair estimate for $\eta_2$:
\begin{equation}
\begin{gathered}
\label{eq:eta2_approx}
	\eta_2=1+\frac{\sqrt{2(5-2\sqrt2)}-5\sqrt2}{4+6\sqrt2} \left( \frac{\gm-1}{\gm} \right)^{0.3} \\
	\approx 1-0.4\left( \frac{\gm-1}{\gm} \right)^{0.3}.
\end{gathered}
\end{equation}

\section*{Data availability}
The data that supports the findings of this study are available within the article.% [and its supplementary material].

\bibliographystyle{unsrt}
\bibliography{references}

\end{document}